\newcommand{\extended}[1]{}    
\newcommand{\short}[1]{#1}     
\renewcommand{\extended}[1]{#1} 
\renewcommand{\short}[1]{}      

\documentclass[10pt]{article}

\usepackage[utf8]{inputenc}
\usepackage[T1]{fontenc}
\usepackage{url}
\usepackage{natbib}
\usepackage{xspace}
\usepackage{framed}
\usepackage{enumitem}

\usepackage{wojtek15logics,wojtek15other}

\usepackage[dvipsnames]{xcolor}

\newcommand{\cit}[1]{\cite[]{#1}}
\newcommand{\Covid}{COVID-19\xspace}
\newcommand{\nind}{\noindent}
\renewcommand{\dot}{\,.\,}

\definecolor{tucgreen}{RGB}{0,140,79}
\newcommand{\alert}[1]{{\color{tucgreen}#1}}

\newenvironment{goals}{\begin{enumerate2}[label=(\roman*)]\color{Maroon}}{\end{enumerate2}}
\newenvironment{requirements}{\begin{enumerate2}[label=(\arabic*)]\color{Maroon}}{\end{enumerate2}}   
\newenvironment{risks}{\begin{enumerate2}[label=(\alph*)]\color{Maroon}}{\end{enumerate2}}    
\newenvironment{formalize}{\[\color{tucgreen}}{\]}

\title{A Survey of Requirements \\ for COVID-19 Mitigation Strategies \medskip \\
  Part II: Elicitation of Requirements}
\author{
  Wojciech Jamroga$^{1,2}$
}
\date{\small
  ${}^1$ Interdisciplinary Centre on Security, Reliability and Trust, SnT, \\ University of Luxembourg
  \\
  ${}^2$ Institute of Computer Science, Polish Academy of Sciences, \\ Warsaw, Poland
}

\begin{document}

\maketitle

\begin{abstract}
The COVID-19 pandemic has influenced virtually all aspects of our lives.
Across the world, countries have applied various mitigation strategies, based on social, political, and technological instruments.
We postulate that multi-agent systems can provide a common platform to study (and balance) their essential properties.
We also show how to obtain a comprehensive list of the properties by ``distilling'' them from media snippets.
Finally, we present a preliminary take on their formal specification, using ideas from multi-agent logics.

\medskip
\nind
\end{abstract}


\section{Introduction}

\Covid has influenced virtually all aspects of our lives.
Across the world, countries applied wildly varying mitigation strategies for the epidemic, ranging from minimal intrusion\extended{ in the hope of obtaining ``herd immunity'',} to imposing severe lockdowns\extended{ on the other extreme}.
%
It seems clear at the first glance what all those measures are trying to achieve, and what the criteria of success are. But is it really that clear?
Quoting an oft-repeated phrase, with \Covid we fight \emph{an unprecedented threat to {health} and {economic stability}}~\cit{Brookings-Tech-Stream-CTAs-2020-04-27}.
While fighting it, we must \emph{protect {privacy}, {equality} and {fairness}}~\cit{Nature-Comment-Ethical-guidelines-2020-06-04}
and
\emph{do a {coordinated assessment} of {usefulness}, {effectiveness}, {technological readiness}, {cyber security risks} and threats to {fundamental freedoms} and {human rights}}~\cit{euobserver-Dutch-soap-opera-2020-05-07}.
Taken together, this is hardly a straightforward set of goals and requirements.
Thus, paraphrasing \cit{euobserver-Dutch-soap-opera-2020-05-07}, one may ask:
\begin{center}
\textbf{What problem does a COVID mitigation strategy solve exactly?}
\end{center}

Even a quick survey of news articles, manifestos, and research papers published since the beginning of the pandemic reveals a diverse landscape of postulates and opinions. Some authors focus on medical goals, some on technological requirements; others are concerned by the economic, social, or political impact of a containment strategy.
The actual stance is often related to the background of the author (in case of a researcher) or their information sources (in case of a journalist).
Moreover, the authors advocating a particular aspect of the strategy most often neglect all the other aspects.
We propose that the field of multi-agent systems can offer a common platform to study all the relevant properties, due to its interdisciplinary nature~\cit{Weiss99mas,Wooldridge02intromas,Shoham09MAS}, well developed theories of heterogeneous agency~\cit{Bratman87intentions,Cohen90intention,Rao91BDI,Wooldridge00rational,Broersen01BOID}, and a wealth of formal methods for specification and verification~\cit{Dastani10MAS,Shoham09MAS,Jamroga15specificationMAS}.

This still leaves the question of how to gather the \emph{actual} goals and requirements for a \Covid mitigation strategy. One way to achieve it is to look at what is considered relevant by the general public, and referred to in the media.
To this end, we collected a number of news quotes on the topic, ordered them thematically and with respect to the type of concern, and presented in~\cit{Jamroga20newsclips}.
Here, we take the news clips from~\cit{Jamroga20newsclips}, and distill a comprehensive list of goals, requirements, and most relevant risk. 
The list is presented in Section~\ref{sec:requirements}. 
In Section~\ref{sec:formalspec}, we make the first step towards a formalization of the properties by formulas of multi-agent logics. We conclude in Section~\ref{sec:conclusions}.

Besides potential input to the design of anti-\Covid strategies, the main contribution of this paper is methodological: we demonstrate how to obtain a comprehensive and relatively unbiased specification of properties for complex MAS by searching for hints in the public space.


\section{Extracting Goals and Requirements from News Clips}
\label{sec:requirements}

Specification of properties is probably the most neglected part of formal verification for MAS.
\extended{The research on formal verification usually concentrates on defining the decision problem, establishing its theoretical properties, and designing algorithms that solve the problem at an abstract level~\cite{Dastani10MAS}. Fortunately, the algorithms are more and more often implemented in the form of a publicly available model-checker~\cit{Alur00mocha,Lomuscio17mcmas,Behrmann04uppaal-tutorial,Kant15ltsmin,Kurpiewski19stv-demo}.}
The tools come with examples of how to model the behavior of a system\short{~\cit{Alur00mocha,Lomuscio17mcmas,Behrmann04uppaal-tutorial,Kant15ltsmin,Kurpiewski19stv-demo}}, but writing the input formulas is generally considered easy. The big question, however, is: \emph{Where do the formulas come from?}
\extended{In a realistic multi-agent scenario, it is not clear at all.}

Mitigating \Covid illustrates the point well. Research on mitigation measures is typically characterized by: (a) strong focus on the native domain of the authors, and (b) focus on the details, rather than the general picture. In order to avoid ``overlooking the forest for the trees,'' we came up with a different methodology.
We looked for relevant phrases{ that appeared} in the media, with no particular method of source selection~\cit{Jamroga20newsclips}.
Then, we extracted the properties,
\extended{and whenever possible generalized statements on specific measures to the mitigation strategy in general.
Finally, we }sorted them thematically, and divided into 3 categories: \emph{goals}, additional \emph{requirements}, and potential \emph{risks and threats}.

While most of the collected snippets focus on digital contact tracing, the relevance of the requirements goes clearly beyond that\extended{, and applies to all the aspects of this epidemic, as well as the ones that may happen in the future}.



\subsection{Epidemiological\extended{ and Health-Related} Concerns}
\label{sec:epid}

\Covid is first and foremost a threat to people's health and lives.
Accordingly, we begin with requirements related to this aspect\extended{ of mitigation strategies}.

\subsubsection{Epidemiological Goals}

The goal of the mitigation strategy in general, and digital measures in particular, is to:
%
\begin{goals}
\item
 provide an \emph{epidemic response}~\cit{Brookings-Tech-Stream-CTAs-2020-04-27}
\item\label{it:under-control}
 \emph{bring the pandemic under control}~\cit{Nature-Comment-Ethical-guidelines-2020-06-04}
\item\label{it:slow-spread}
 \emph{slow the spread of the virus}~\cit{Top10VPN-Digital-Rights-2020-06-10,NCSC-nhs-explainer-2020-05-04,Brookings-Tech-Stream-CTAs-2020-04-27,Telecoms-Unlike-France-2020-04-27,helsenorge-Together-2020-04-28,Guardian-democracies-2020-06-16}
\item\label{it:prevent-deaths}
 \emph{prevent deaths}~\cit{RTL-LU-Lockdowns-averted-2020-06-09}
\item
 \emph{reduce the reproduction rate} of the virus\extended{, i.e., how many people are infected by someone with the virus}~\cit{RTL-LU-Lockdowns-averted-2020-06-09}.
\end{goals}

  \nind
  The specific goals of digital measures are to:
  \begin{goals}
  \item
   \emph{trace the spread of the virus} and \emph{identify \extended{dangerous }Covid-19 clusters}~\cit{Guardian-democracies-2020-06-16}
  \item
   \emph{find potential new infections}~\cit{WashingtonPost-Most-Americans-2020-04-29}
  \item
   \emph{register contacts between potential carriers\extended{ and those who might be infected}}~\cit{Guardian-democracies-2020-06-16}
  \item
   \emph{deter people from breaking quarantine}~\cit{BBC-News-Why-Indias-2020-05-15}
  \end{goals}

\nind
Requirements:
\begin{requirements}
\item
 The efforts must meet \emph{public health needs} best~\cit{Brookings-Tech-Stream-CTAs-2020-04-27,Guardian-democracies-2020-06-16}.
\item
 Digital measures should\short{ \emph{complement} traditional \short{ones}\extended{forms of mitigation}~\cit{Brookings-Tech-Stream-CTAs-2020-04-27,WashingtonPost-Most-Americans-2020-04-29}}\extended{ be a \emph{component of the epidemic response}~\cit{Brookings-Tech-Stream-CTAs-2020-04-27}, and \emph{enhance traditional forms of contact tracing}~\cit{WashingtonPost-Most-Americans-2020-04-29}}
\item
 They should be designed to \emph{help the health authorities}~\cit{helsenorge-Together-2020-04-28}.
\end{requirements}

\subsubsection{Effectiveness of Epidemic Response}
\label{sec:effectiveness}

Requirements:
\begin{requirements}
\item
 The strategy should be \emph{effective}~\cit{Brookings-Tech-Stream-CTAs-2020-04-27,euobserver-Dutch-soap-opera-2020-05-07}
\item\label{it:difference}
 It should \emph{make a difference}~\cit{Wired-Just-how-2020-05-12}.
\end{requirements}

\nind
Risks and threats: 
\begin{risks}
\item
 \emph{Inaccurate detection} of carriers and infected people\extended{ due to the limitations of the technology and the underlying model of human interaction}~\cit{Brookings-Tech-Stream-CTAs-2020-04-27}
\item
 Specifically, this may \emph{adversely impact \short{easing}\extended{relaxation} of lockdowns}~\cit{Top10VPN-Digital-Rights-2020-06-10}
\item
 Misguided \emph{assurance} that going out is safe~\cit{Brookings-Tech-Stream-CTAs-2020-04-27}.
\end{risks}

\subsubsection{Information Flow\short{ Requirements}\extended{ to Counter the Epidemic}}
\label{sec:infoflow}

The strategy should\extended{ support rapid identification and notification of the most concerned. That is, it should} allow:
\begin{requirements}
\item\label{it:identify-exposed}
 to \emph{identify people who might have been exposed to the virus}~\cit{Nature-Can-they-slow-2020-05-19}
\item\label{it:alert}
 to \emph{alert those people}~\cit{Nature-Comment-Ethical-guidelines-2020-06-04,helsenorge-Together-2020-04-28,WashingtonPost-Most-Americans-2020-04-29,POLITICO-States-struggle-2020-05-17}.
\item\label{it:alert-rapid}
 The identification and notification must be \emph{rapid}~\cit{Nature-Can-they-slow-2020-05-19,Nature-Comment-Ethical-guidelines-2020-06-04}.
\end{requirements}

\subsubsection{Monitoring\extended{ Pandemic and Containment Strategy}}
\label{sec:monitoring}

The containment strategy should enable:
\begin{requirements}
\item\label{it:monitoring-1}
 \emph{monitoring the state of the pandemic}, e.g., the outbreaks and the spread of the virus~\cit{POLITICO-States-struggle-2020-05-17,ScientificAmerican-AntibodyTests-2020-06-17}
\item
 \emph{monitoring the behavior of people}, in particular if they are following the rules~\cit{POLITICO-Polands-2020-04-02}
\item
 to \emph{monitor the effectiveness of the strategy}~\cit{Telecoms-UK-2020-04-28}.
\end{requirements}

\subsubsection{Tradeoffs}

There are \emph{tradeoffs} between {effective containment of the epidemic} and other concerns\extended{, such as \emph{privacy} and protection of \emph{fundamental freedoms}}~\cit{Register-UK-finds-2020-05-05,BBC-News-Why-Indias-2020-05-15,POLITICO-States-struggle-2020-05-17,Guardian-democracies-2020-06-16}.
E.g., effective monitoring is often at odds with privacy~\cit{Telecoms-UK-2020-04-28}.
The strategy should
\begin{requirements}
\item
 \emph{strike the right balance} between different concerns~\cit{Guardian-democracies-2020-06-16}.
\end{requirements}
\extended{We will see more tradeoff-related requirements in the subsequent sections.}




\subsection{Economic and Social Impact}
\label{sec:econ+soc}

Most measures to contain the epidemic are predominantly social (cf.~lockdown), and have strong social and economic impact.

\subsubsection{Economic Stability}

The containment strategy should:
\begin{requirements}
\item
  minimize the \emph{cost to local economies} and the negative impact on \emph{economic growth}~\cit{Brookings-Tech-Stream-CTAs-2020-04-27,RTL-LU-Lockdowns-averted-2020-06-09}
\item
 allow for \emph{return to normal economy and society} and make resumption of economic and social activities \emph{safer}~\cit{WashingtonPost-Most-Americans-2020-04-29,Guardian-Covidsafe-2020-05-23}.
\end{requirements}

\subsubsection{Social and Political Impact}

The containment strategy (and digital measures in particular) should:
\begin{requirements}
\item
 \emph{ease lockdowns} and  \emph{home confinement}~\cit{Brookings-Tech-Stream-CTAs-2020-04-27,euobserver-Dutch-soap-opera-2020-05-07,Nature-Can-they-slow-2020-05-19,Guardian-Covidsafe-2020-05-23}
\item
 minimize adverse impact on \emph{social relationships} and \emph{personal well-being}~\cit{Brookings-Tech-Stream-CTAs-2020-04-27}
\item
 \emph{prohibit economic and social discrimination} on the basis of information and technology being part of the strategy~\cit{Brookings-Tech-Stream-CTAs-2020-04-27}
\item
 \emph{protect the communities} that can be harmed by the collection and exploitation of personal data~\cit{Brookings-Tech-Stream-CTAs-2020-04-27}.
\end{requirements}

\extended{
  \nind
  Detailed requirements:
  \begin{requirements}
  \item
   Surveillance technologies should not become \emph{compulsory for public and social engagements}, with unaffected individuals \emph{restricted from participating in social and economic activities}~\cit{Brookings-Tech-Stream-CTAs-2020-04-27}.
  \end{requirements}
}

\nind
Risks and threats:
\begin{risks}
\item
 \emph{Little knowledge} about social impact of the measures~\cit{MIT-covid-tracing-2020-05-07}
\item
 \emph{Discrimination} and creation of \emph{social divides}~\cit{Brookings-Tech-Stream-CTAs-2020-04-27,MatrixChambers-Legal-Advice-2020-05-03}
\item
 \emph{Disinformation} and \emph{information abuse}~\cit{Brookings-Tech-Stream-CTAs-2020-04-27,Top10VPN-Digital-Rights-2020-06-10}
\item
 Providing a \emph{false sense of security}~\cit{Brookings-Tech-Stream-CTAs-2020-04-27}
\item
 \emph{Political manipulation}, creating \emph{social unrest}, and \emph{dishonest competition} by false reports of coronavirus~\cit{Brookings-Tech-Stream-CTAs-2020-04-27}
\item
 Too much political influence of IT companies on \emph{the decisions of sovereign democratic countries}~\cit{Guardian-democracies-2020-06-16}.
\end{risks}


\subsubsection{Costs, Human Resources, Logistics}\label{sec:costs}
Requirements:
\begin{requirements}
\item
 The \emph{financial cost} of the measures should be minimized~\cit{Guardian-UK-abandons-2020-06-18}
\item
 Minimization of the involved \emph{human resources}~\cit{POLITICO-Polands-2020-04-02,Brookings-Tech-Stream-CTAs-2020-04-27}
\item
 \emph{Timeliness}~\cit{Guardian-UK-abandons-2020-06-18}
\item
 \emph{Coordination} between different institutions and authorities\cit{Politico-Google-Apple-2020-06-10,EuropeanCommission-interoperability-2020-06-16}, including the establishment of \emph{common standards}~\cit{Politico-Google-Apple-2020-06-10}.
\end{requirements}



\subsection{Ethical and Legal Aspects}

In this section, we look at requirements that aim at the long-term robustness and resilience of the social structure.

\subsubsection{Ethical and Legal Requirements}
\begin{requirements}
\item
 The mitigation strategy must be \emph{ethically justifiable}~\cit{Nature-Comment-Ethical-guidelines-2020-06-04}
\item
 The measures should be \emph{necessary}, \emph{proportionate}, \emph{legitimate}, \emph{just}, \emph{scientifically valid}, and \emph{time-bound}~\cit{Nature-Comment-Ethical-guidelines-2020-06-04,Top10VPN-Digital-Rights-2020-06-10,Guardian-Norway-2020-06-15,POLITICO-Polands-2020-04-02,MatrixChambers-Legal-Advice-2020-05-03}
\item
 They should not be \emph{invasive}~\cit{BBC-News-Why-Indias-2020-05-15} and must not be done at the expense of \emph{individual civil rights}~\cit{Telecoms-Unlike-France-2020-04-27,MIT-covid-tracing-2020-05-07,MatrixChambers-Legal-Advice-2020-05-03}
\item
 Means of protection should be \emph{available to anyone}~\cit{Nature-Comment-Ethical-guidelines-2020-06-04}
\item
 They should be \emph{voluntary}~\cit{MIT-covid-tracing-2020-05-07,NCSC-nhs-explainer-2020-05-04}
\item
 The measures must \emph{comply with legal regulations}~\cit{MatrixChambers-Legal-Advice-2020-05-03,Register-UK-finds-2020-05-05,Gizmodo-UKs-Contact-Tracing-2020-05-13}
\item
 \emph{Implementation} and \emph{impact} must also be considered~\cit{Nature-Comment-Ethical-guidelines-2020-06-04,Top10VPN-Digital-Rights-2020-06-10}
\item
 \emph{Impact assessment} should be \emph{conducted} and \emph{made public}~\cit{MatrixChambers-Legal-Advice-2020-05-03}.
\end{requirements}

\subsubsection{Risks and Threats}
\begin{risks}
\item
 Serious and long-lasting \emph{harms to fundamental rights and freedoms}~\cit{Nature-Comment-Ethical-guidelines-2020-06-04}
\item
 Costs of \emph{not devoting resources to something else}~\cit{Nature-Comment-Ethical-guidelines-2020-06-04}
\extended{
  \item
   Measures designed and implemented \emph{without adequate scrutiny}~\cit{Top10VPN-Digital-Rights-2020-06-10}
}
\item
 Measures that support \emph{extensive physical surveillance}~\cit{Top10VPN-Digital-Rights-2020-06-10}
\item
 \emph{Mandatory} use of digital measures, \emph{collecting sensitive information}, \emph{sharing the data} with the government~\cit{BBC-News-Why-Indias-2020-05-15,Nature-Can-they-slow-2020-05-19}
\item
 \emph{Censorship practices}\extended{ to silence critics and control the flow of information}~\cit{Top10VPN-Digital-Rights-2020-06-10}.
\end{risks}





\subsection{Privacy and Data Protection}
\label{sec:privacy}

Privacy-related issues for \Covid mitigation strategies have triggered heated discussion, and \extended{at some point }gained much media coverage.
\extended{%
  This is understandable, since privacy and data protection is an important aspect of medical information flow, even in ordinary times.
  Moreover, the IT measures against \Covid are usually designed by computer scientists and specialists, for whom security requirements are relatively easy to identify and understand.
}

\subsubsection{General Privacy Requirements}
\label{sec:general-priv}

\begin{requirements}
\item
 The strategy should be designed with \emph{privacy} and \emph{information security} in mind~\cit{Brookings-Tech-Stream-CTAs-2020-04-27,WashingtonPost-Most-Americans-2020-04-29,MIT-covid-tracing-2020-05-07}
\extended{
  \item
   It should \emph{mitigate privacy concerns} inherent in a \emph{technological approach}~\cit{Brookings-Tech-Stream-CTAs-2020-04-27}
}
\item\label{it:anonymous}
 It should be \emph{anonymous} under data protection laws, i.e., it cannot \emph{lead to the identification of an individual}~\cit{Wired-Just-how-2020-05-12}
\item
 The \emph{information} about users should be \emph{protected at all times}~\cit{NCSC-nhs-explainer-2020-05-04}
\item
 The design should include recommendations for \emph{how back-end systems should be secured}, and identify \emph{vulnerabilities} as well as \emph{unintended consequences}~\cit{Brookings-Tech-Stream-CTAs-2020-04-27}.
\end{requirements}

\nind
Risks and Threats:
\begin{risks}
\item
 Lack of \emph{clear privacy policies}~\cit{Top10VPN-Digital-Rights-2020-06-10,Politico-Privacy-fears-2020-06-04,MIT-covid-tracing-2020-05-07}
\item
 \emph{Exploitation of personal information} by authorities or third parties~\cit{Politico-Privacy-fears-2020-06-04,Top10VPN-Digital-Rights-2020-06-10,BBCNews-Alarm-Kuwait-2020-06-16}\extended{, in particular \emph{live or near-live tracking of users' locations} and \emph{linking sensitive personal information to an individual}~\cit{BBCNews-Alarm-Kuwait-2020-06-16}}
\item
 \emph{Linking different datasets} at some point in the future~\cit{Gizmodo-UKs-Contact-Tracing-2020-05-13}
\item
 Alerts can be \emph{too revealing}~\cit{BBC-Coronavirus-privacy-2020-03-05}
\item
 It may be possible to work out \emph{who is associating with whom}~\cit{Register-UK-finds-2020-05-05}.
\end{risks}

\subsubsection{Data Protection and Potential Misuse of Data}
\label{sec:data}

Here, the key question is:
\emph{What data is collected} and \emph{who is it shared with}?~\cit{MIT-covid-tracing-2020-05-07,Brookings-Tech-Stream-CTAs-2020-04-27}
This leads to the following requirements:
\begin{requirements}
\item
 Clear and reasonable \emph{limits on the data collection types}~\cit{Politico-Google-Apple-2020-06-10,BBC-News-Why-Indias-2020-05-15,NCSC-nhs-explainer-2020-05-04,Brookings-Tech-Stream-CTAs-2020-04-27,WashingtonPost-Most-Americans-2020-04-29}
\item
 Limitations on \emph{how the data is used}~\cit{MIT-covid-tracing-2020-05-07}
\item
 In particular, the data is to be \emph{used strictly for disease control} and not \emph{shared with law enforcement agencies}~\cit{BBC-News-Why-Indias-2020-05-15,Guardian-Covidsafe-2020-05-23}
\item
 Less \emph{state access} and \emph{control} over \emph{user data}~\cit{Telecoms-Unlike-France-2020-04-27}
\item
 Data collection should be \emph{minimized}~\cit{MIT-covid-tracing-2020-05-07} and based on \emph{informed consent} of the participants~\cit{Guardian-democracies-2020-06-16}
\item\label{it:access1}
 Giving access to one's data should be \emph{voluntary}~\cit{MIT-covid-tracing-2020-05-07}
\item
 One should be able to \emph{delete their personal information}\extended{ at any time}~\cit{helsenorge-Together-2020-04-28,Register-UK-finds-2020-05-05}
\item
 One should have the \emph{right to access their own data}~\cit{helsenorge-Together-2020-04-28,Register-UK-finds-2020-05-05}
\item
 For digital measures, the user should be able to \emph{remove the software} and \emph{disable more invasive features}~\cit{helsenorge-Together-2020-04-28}.
\end{requirements}


\nind
Risks and threats:
\begin{risks}
\item
 Data storage that can be \emph{hacked} and \emph{exploited}~\cit{Telecoms-UK-2020-04-28,Nature-Can-they-slow-2020-05-19,Top10VPN-Digital-Rights-2020-06-10}
\item
 \emph{Data breaches} due to insider threats~\cit{Politico-Privacy-fears-2020-06-04}
\item
 \emph{Function creep} and \emph{state surveillance}~\cit{Nature-Can-they-slow-2020-05-19}
\item
 \emph{Sharing data} across agencies or \emph{selling} to a third party~\cit{Politico-Privacy-fears-2020-06-04,Top10VPN-Digital-Rights-2020-06-10}
\item
 Integration with \emph{commercial services}~\cit{Top10VPN-Digital-Rights-2020-06-10}.
\end{risks}

\subsubsection{Sunsetting, Safeguards, and Monitoring}

Requirements:
\begin{requirements}
\item
 \extended{Sunsetting: the m}\short{M}easures should be \emph{terminated} as soon as possible~\cit{POLITICO-Polands-2020-04-02,Brookings-Tech-Stream-CTAs-2020-04-27,helsenorge-Together-2020-04-28}
\item
 Data should be eventually or even periodically \emph{destroyed}~\cit{POLITICO-Polands-2020-04-02,MIT-covid-tracing-2020-05-07,helsenorge-Together-2020-04-28,Register-UK-finds-2020-05-05,Brookings-Tech-Stream-CTAs-2020-04-27,WashingtonPost-Most-Americans-2020-04-29}, in particular \emph{when it is no longer needed to help manage the spread of coronavirus}~\cit{NCSC-nhs-explainer-2020-05-04}
\item
 \emph{Transparency} of data collection~\cit{MIT-covid-tracing-2020-05-07}
\item
 There should be clear \emph{policies to prevent abuse}~\cit{MIT-covid-tracing-2020-05-07}
\item
 Privacy must be backed up with clear lines of \emph{accountability} and processes for \emph{evaluation} and \emph{monitoring}~\cit{Gizmodo-UKs-Contact-Tracing-2020-05-13}
\item
 \emph{Judicial oversight} must be provided~\cit{Brookings-Tech-Stream-CTAs-2020-04-27}
\item
 Safeguards should be backed by \emph{an independent figure}~\cit{POLITICO-Polands-2020-04-02}.
\end{requirements}

\nind
Risks and threats:
\begin{risks}
\item
 Surveillance might \emph{continue to be used} after the \short{pandemic}\extended{threat of the coronavirus recedes}~\cit{BBCNews-Alarm-Kuwait-2020-06-16}
\item
 Data can \emph{stay with the government} longer than necessary~\cit{POLITICO-Polands-2020-04-02}.
\end{risks}

\subsubsection{Impact of Privacy}

Requirements:
\begin{requirements}
\item
 People must \emph{get the information they need} to protect themselves and others~\cit{BBC-Coronavirus-privacy-2020-03-05}
\item
 There must be protections against \emph{economic and social discrimination} based on \emph{information} and \emph{technology} designed to fight the pandemic\extended{, in particular with respect to communities vulnerable to \emph{collection and exploitation of personal data}}~\cit{Brookings-Tech-Stream-CTAs-2020-04-27}
\item
 Information should be used in such a way that people who fear being judged will not \emph{put other people in danger}~\cit{BBC-Coronavirus-privacy-2020-03-05}.
\end{requirements}

\nind
Risks and threats:
\begin{risks}
\item
 \emph{Fear of social stigma}~\cit{BBC-Coronavirus-privacy-2020-03-05}
\item
 Online \emph{judgement} and \emph{ridicule}~\cit{BBC-Coronavirus-privacy-2020-03-05}.
\end{risks}

\extended{
  \subsubsection{Privacy vs.~Epidemiological Efficiency}
  \label{sec:priv-vs-epid}
  \label{sec:priv-support-epid}

  There is a tradeoff between protecting privacy vs.~collecting and processing all the information that can be useful in fighting the epidemic:
  \begin{itemize2}
  \item
   Privacy hinders \emph{making the best possible use of the data}, including \emph{analysis of the population}, \emph{contact matching}, \emph{modeling the network of contacts}, enabling \emph{epidemiological insights} such as \emph{revealing clusters} and \emph{superspreaders}, and providing \emph{advice to people}~\cit{Register-UK-finds-2020-05-05,Nature-Can-they-slow-2020-05-19,Guardian-Covidsafe-2020-05-23}
  \item
   Privacy-preserving solutions put users in \emph{more control of their information} and require \emph{no intervention from a third party}~\cit{Register-UK-finds-2020-05-05}.
  \end{itemize2}

  \nind
  The relationship is not simply antagonistic, though:
  \begin{itemize2}
  \item
   Privacy is instrumental in building \emph{trust}. Conversely, lack of privacy undermines trust, and may \emph{hinder the epidemiological, economic, and social effects of the mitigation activities}~\cit{Politico-Privacy-fears-2020-06-04}.
  \end{itemize2}
}

\subsubsection{Reasonable Privacy}
\short{\label{sec:priv-support-epid}}

While it might be necessary to waive users’ privacy in the short term in order to contain the epidemic, one must look for mechanisms such that
\begin{requirements}
\item
 \emph{exploiting the risks would require significant effort by the attackers for minimal reward}~\cit{Nature-Can-they-slow-2020-05-19}.
\end{requirements}
\short{Moreover, there is a tradeoff between privacy and collecting the information that can be useful in fighting the epidemic~\cit{Register-UK-finds-2020-05-05,Nature-Can-they-slow-2020-05-19,Guardian-Covidsafe-2020-05-23}.
On the other hand, lack of privacy \emph{undermines trust}, and may \emph{hinder the effectiveness of the mitigation activities}~\cit{Politico-Privacy-fears-2020-06-04}. }




\subsection{User-Related Aspects}

The measures must be adopted\extended{ and followed by the people,} in order to make them effective.

\subsubsection{User Incentives}
\label{sec:userinc}
Goals:
\begin{goals}
\item
 \emph{High acceptance rate} for the mitigation measures~\cit{WashingtonPost-Most-Americans-2020-04-29}.
\item
 \emph{Creating incentives} and overcoming \emph{incentive problems} for individual people to adopt the strategy~\cit{Brookings-Tech-Stream-CTAs-2020-04-27}
\end{goals}

\nind
Risks and threats:
\begin{risks}
\item
 Lack of \emph{immediate benefits} for the participants~\cit{Brookings-Tech-Stream-CTAs-2020-04-27}
\item
 Perceived \emph{privacy} and \emph{security risks}~\cit{WashingtonPost-Most-Americans-2020-04-29}
\item
 Some measures can \emph{divert attention from more important measures}, and \emph{make people less alert}~\cit{Panoptykon-ProteGo-Safe-2020-05-06}
\item
 Creating \emph{false sense of security} from the pandemic~\cit{ScientificAmerican-AntibodyTests-2020-06-17}.
\end{risks}

\extended{
  \nind
  Countermeasures:
  \begin{risks}
  \item
   Pointing out \emph{indirect benefits} (e.g., opening of the schools and businesses, reviving the national economy)~\cit{Brookings-Tech-Stream-CTAs-2020-04-27}
  \item
   Reliance on \emph{personal responsibility}~\cit{euobserver-Dutch-soap-opera-2020-05-07}.
  \end{risks}
}

%

\subsubsection{Adoption and Its Impact}

Requirements:
\begin{requirements}
\item
 \emph{Enough people} should \emph{download} and \emph{use} the app to make it \emph{effective}~\cit{WashingtonPost-Most-Americans-2020-04-29,MIT-covid-tracing-2020-05-07,Nature-Can-they-slow-2020-05-19,LeMonde-StopCovid-2020-06-10}.
 Note: this requirement is \emph{graded} rather than binary~\cite{MIT-Technology-coronavirus-apps-2020-06-05,Hinch20covid-uptake}.
\end{requirements}


\nind
Risks and threats:
\begin{risks}
\item
 Lack of users' \emph{trust}~\cit{Wired-Just-how-2020-05-12,Politico-Privacy-fears-2020-06-04}\extended{, see also the connection between privacy and trust in Section~\ref{sec:priv-support-epid}}
\item
 Lack of \emph{social knowledge} and \emph{empathy} by the authorities~\cit{POLITICO-States-struggle-2020-05-17}.
\end{risks}

\subsection{Technological Aspects}
General requirements:
\begin{requirements}
\item
 The concrete measures and tools must be \emph{operational}~\cit{Register-UK-finds-2020-05-05,POLITICO-Polands-2020-04-02}
\item
 In particular, they should be \emph{compatible} with their environment of implementation~\cit{Gizmodo-UKs-Contact-Tracing-2020-05-13}
\item
 Design and implementation should be \emph{transparent}~\cit{MIT-covid-tracing-2020-05-07,SDZ-Corona-App-2020-05-06}.
\end{requirements}

\nind
Specific requirements for digital measures:
\begin{requirements}
\item
 They should be \emph{compatible with most available devices}~\cit{Gizmodo-UKs-Contact-Tracing-2020-05-13}
\item
 Reasonable \emph{use of battery}~\cit{Gizmodo-UKs-Contact-Tracing-2020-05-13}
\item
 \emph{Usable interface}~\cit{Gizmodo-UKs-Contact-Tracing-2020-05-13}
\item
 \emph{Accurate measurements} of how close two devices are~\cit{Nature-Can-they-slow-2020-05-19}
\item
 \emph{Cross-border interoperability}~\cit{Cybernetica-Proposes-2020-05-06}
\item
 Possibility to \emph{verify the code} by the public and experts~\cit{SDZ-Corona-App-2020-05-06}.
\end{requirements}




\subsection{Evaluation and Learning for the Future}

\Covid mitigation activities should be rigorously assessed. Moreover, their outcomes should be used to extend our knowledge\extended{ about the pandemic}, and better defend ourselves in the future.
The main goal here is:
\begin{goals}
\item
 to use the collected data in order to \emph{develop efficient infection control measures} and \emph{gain insight into the effect of changes to the measures for fighting the virus}~\cit{helsenorge-Together-2020-04-28,Register-UK-finds-2020-05-05}.
\end{goals}

\nind
Requirements:
\begin{requirements}
\item
 A \emph{review} and \emph{exit strategy} should be defined~\cit{Nature-Comment-Ethical-guidelines-2020-06-04}
\item
 Before implementing the measures, an \emph{institutional assessment} is needed of their \emph{usefulness, effectiveness, technological readiness, cyber-security risks and threats to fundamental freedoms and human rights}~\cit{euobserver-Dutch-soap-opera-2020-05-07}
\item
 After the pandemic, there must be \emph{the society's assessment} whether the strategy has been effective and appropriate~\cit{BBC-Coronavirus-privacy-2020-03-05}
\item
 The \emph{assessments} should be conducted \emph{by an independent body} at \emph{regular intervals}~\cit{Nature-Comment-Ethical-guidelines-2020-06-04}.
\end{requirements}



\section{Towards Formal Specification}
\label{sec:formalspec}

Here, we briefly show how the requirements presented in Section~\ref{sec:requirements} can be rewritten in a more formal way.
To this end, we use \emph{modal logics for distributed and multi-agent systems} that have been in constant development for over 40 years~\cit{Emerson90temporal,Fagin95knowledge,Wooldridge00rational,Broersen01BOID,Alur02ATL,Goranko15stratmas}.
Note that the following specifications are only \emph{semi}-formal, as we do not fix the models nor give the precise semantics of the logical operators\extended{ and atomic predicates}.
We leave that step for the future work.

\subsection{Temporal and Epistemic Properties}
\label{sec:tempoepist}

The simplest kind of requirements are those that refer to achievement or maintenance of a particular state of affairs.
Typically, they can be expressed by formulas of the branching-time logic \CTLs~\cit{Emerson90temporal}, with path quantifiers $\Epath$ (\emph{there is a path}), $\Apath$ (\emph{for all paths}), and temporal operators
$\Next$ (\emph{in the next moment}), $\Sometm$ (\emph{sometime from now on}), $\Always$ (\emph{always from now on}), and $\Until$ (\emph{until}).
For example, goal~\ref{it:under-control} in Section~\ref{sec:epid} can be tentatively rewritten as the \CTLs formula\\
\begin{formalize}
\Apath\Sometm\prop{control-pandemic},
\end{formalize}
saying that, for all possible execution paths, \alert{$\prop{control-pandemic}$} must eventually hold.\footnote{
  In fact, a better specification is given by\ \alert{$\Apath\Sometm\Always\prop{control-pandemic}$}, saying that the pandemic is not only brought, but also kept under control from some point on. }
Similarly, goal~\ref{it:slow-spread} can be expressed by formula\ \alert{$\forall n\dot (\prop{R0=n}) \then \Apath\Sometm(\prop{R0<n})$}.
\extended{It is easy to see that the latter requirement is more precise than the former.}
Moreover, goal~\ref{it:prevent-deaths} can be captured by\ \alert{$\Apath\Always(\prop{\#deaths<k})$}, for a reasonably chosen $k$.

The identification and monitoring aspects can be expressed by a combination of \CTLs with epistemic operators $K_a\varphi$ (\emph{``$a$ knows that $\varphi$''}).
For example, the information flow requirement~\ref{it:identify-exposed} in Section~\ref{sec:infoflow} can be transcribed as\ \alert{$\prop{exposed}_i \then \Apath\Sometm K_a\prop{exposed}_i$}, where $a$ is the name of the agent (or authority) supposed to identify the vulnerable people.
A more faithful transcription can be obtained using the past-time operator $\Sometm^{-1}$ (\emph{sometime in the past})~\cit{Laroussinie95hierarchy} with
\begin{formalize}
exp_i \then \Apath\Sometm K_a exp_i,\quad\textrm{\color{black} where }exp_i \equiv \Sometm^{-1}\prop{exposed}_i,
\end{formalize}
saying that if \alert{$\prop{exposed}_i$} held at some point\extended{ in the past}, then $a$ will eventually know about it.
Likewise, the information flow requirement~\ref{it:alert} can be captured by\ \alert{$K_a exp_i \then \Apath\Sometm K_i exp_i$}.
Similar\extended{ temporal-epistemic} formulas may be used to express some privacy-related requirements in Section~\ref{sec:privacy}\extended{, e.g., \alert{$\forall j\neq i\dot \Apath\Always (\neg K_j (x=i) \land \neg K_j (x\neq i))$} tentatively captures the anonymity of person $i$ wrt.~the database entry represented by $x$, cf.~requirement~\ref{sec:general-priv}.\ref{it:anonymous}}.

\subsection{Strategic Requirements}
\label{sec:strategic}

The\short{ above patterns}\extended{ temporal and epistemic patterns, presented above,} can be refined by replacing path quantifiers $\Apath, \Epath$ with strategic operators $\coop{A}$ of the logic \ATLs~\cit{Alur02ATL,Goranko15stratmas}, where $\coop{A}\varphi$ says that \emph{``the agents in $A$ can bring about $\varphi$''}.
For example, the information flow requirement~\ref{sec:infoflow}.\ref{it:alert} can be rewritten as
\begin{formalize}
K_a (\Sometm^{-1}\prop{exposed}_i) \then \coop{a}\Sometm \coop{i}\Sometm K_i(\Sometm^{-1}\prop{exposed}_i),
\end{formalize}
saying that if the health authority $a$ knows that $i$ was exposed, then $a$ can provide $i$ with the information sufficient to realize that.
Strategic operators are also useful for the monitoring requirements in Section~\ref{sec:monitoring}, e.g.,
\alert{$\coop{a}\Always (K_a\prop{outbreak} \lor K_a\neg\prop{outbreak})$} can be used for requirement~\ref{sec:monitoring}.\ref{it:monitoring-1}.
\extended{The same applies to the access control properties in Section~\ref{sec:data}, e.g., requirement~\ref{it:access1} can be formalized by formula \alert{$\forall d\in data(i)\, \forall j\neq i\dot \coop{i}\Sometm\prop{access(j,d)} \land \coop{i}\Always\neg\prop{access(j,d)}$}.}

For some requirements, this should be combined with bounds imposed on the execution time~\cit{Penczek06advances,Knapik19timedATL}, mental complexity~\cit{Jamroga19natstrat-aij}, and/or resources needed to accomplish the tasks~\short{\cit{Alechina10atl-bounded,Bulling10rtl}}\extended{\cit{Alechina10atl-bounded,Bulling10rtl,Alechina17resourceboundedATL-mcheck}}.
For example, the notification requirement~\ref{sec:infoflow}.\ref{it:alert} can be refined as:
\begin{formalize}
\coop{a}\Sometm^{t\le 10} \coop{i}^{compl\le 5}\Sometm K_i(\Sometm^{-1}\prop{exposed}_i),
\end{formalize}
based on the assumption that $a$ should notify $i$ in at most $10$ time units, and $i$ has a strategy of complexity at most $5$ to infer the relevant knowledge from the notification.

\extended{
  To reason explicitly about the outcome of anti-\Covid measures, we may need model update operators $(supp\ a:\sigma)$ saying \emph{``suppose that agent $a$ plays strategy $\sigma$''}~\cit{Bulling08atlp-amai}, and similar to strategy binding in Strategy Logic~\short{\cit{Mogavero10stratLogic}}\extended{\cit{Mogavero10stratLogic} and \ATL with Strategy Contexts~\cit{Brihaye09strategycontexts}}.
  Then, the effectiveness requirement~\ref{it:difference} of Section~\ref{sec:effectiveness} for mitigation strategy $\mathcal{S}$ can be written as\\
  \begin{formalize}
  \neg\psi \land (supp\ a:\mathcal{S})\psi,
  \end{formalize}
  where $\psi$ can be any of the properties of epidemic response, formalized in Section~\ref{sec:tempoepist}.
}
\nocite{Bulling08atlp-amai,Mogavero10stratLogic,Brihaye09strategycontexts,Guelev11atl-distrknowldge}

\subsection{Probabilistic Extensions}

Many events have probabilistic execution, e.g., actions may fail with some (typically small) probability. Scenarios with probabilistic events can be modeled by variants of Markov decision processes, and their properties can be specified by a probabilistic variant of \CTLs~\cit{Baier97probabilisticMcheck} or \ATLs~\cit{Chen13prismgames}.
For instance, formula
\begin{formalize}
\coop{a}^{P\ge 0.99}\Sometm^{t\le 10} \coop{i}^{compl\le 5}\Sometm K_i(\Sometm^{-1}\prop{exposed}_i),
\end{formalize}
refines the previous specification by demanding that the authority can successfully notify $i$ with probability at least $99\%$.

\subsection{Meta-Properties}

The requirements presented in Section~\ref{sec:requirements}, and formalized above, refer to the ``correctness'' of a given mitigation strategy.
Two meta-properties, well known in computer science, can be also useful in case of the present scenario, namely \emph{diagnosability} and \emph{resilience}~\cit{Ezekiel17diagnosability}.
Given a correctness requirement $\Phi$ and a responsible agent $a$, those can be expressed by the following templates:
\begin{eqnarray*}
\alert{\Apath\Always(\neg\Phi \then \coop{a}\Sometm K_a\neg\Phi)} && \textrm{\emph{(diagnosability)}} \\
\alert{\Apath\Always(\neg\Phi \then \coop{a}\Sometm\Phi)} && \textrm{\emph{(resilience)}}
\end{eqnarray*}
The templates can be used e.g.~for monitoring-type requirements.

\subsection{Towards Formal Analysis}

Ideally, one would like to automatically evaluate \Covid strategies with respect to the requirements, and choose the best one.
In the future, we plan to use model checking tools, such as MCMAS~\cit{Lomuscio17mcmas}, Uppaal~\cit{Behrmann04uppaal-tutorial}, or PRISM~\cit{Kwiatkowska02prism}, to formally verify our formulas over micro-level models created to simulate and predict the progress of the pandemic~\cit{Ferguson20impactCovid,Adamik20estimationCOVID,Bock20mitigation,McCabe20modellingICU}.
As we already pointed out, different requirements may be in partial conflict. Thus, selecting an optimal mitigation strategy may require solving a multicriterial optimization problem~\cit{Zionts81multicriterial,Collette04multiobjective,Radulescu20multiobjective-survey}, e.g., by identifying the Pareto frontier and choosing a criterion to select a point on the frontier.


\section{Conclusions}
\label{sec:conclusions}

In this paper, we make the first step towards a systematic analysis of strategies for effective and trustworthy mitigation of the current pandemic.
The strategies may incorporate medical, social, economic, as well as technological measures. Consequently, there is a large number of medical, social, economic, and technological requirements that must be taken into account\extended{ when deciding which strategy to adopt}.
For computer scientists, the latter kind of requirements is most natural, which is exactly the pitfall that computer scientists must avoid.
The goals\extended{ (and acceptability criteria) for a mitigation strategy} are much more diverse, and we must consciously choose a solution that satisfies the multiple criteria to a reasonable degree.
We suggest that formal methods for MAS provide an excellent framework for that.
We also propose a methodology to collect preliminary requirements while avoiding the usual bias of research papers.

\para{Acknowledgments.}
The author acknowledges the support of the Luxembourg National Research Fund (FNR) under the COVID-19 project SmartExit,
and the support of the National Centre for Research and Development Poland (NCBR) and the Luxembourg National Research Fund (FNR), under the PolLux/CORE project STV (POLLUX-VII/1/2019).


\nocite{BBC-Coronavirus-privacy-2020-03-05}
\nocite{POLITICO-Polands-2020-04-02}
\nocite{Brookings-Tech-Stream-CTAs-2020-04-27}
\nocite{Telecoms-Unlike-France-2020-04-27}
\nocite{Telecoms-UK-2020-04-28}
\nocite{helsenorge-Together-2020-04-28}
\nocite{Tabletowo-Aplikacja-2020-04-29}
\nocite{WashingtonPost-Most-Americans-2020-04-29}
\nocite{MatrixChambers-Legal-Advice-2020-05-03}
\nocite{NCSC-nhs-explainer-2020-05-04}
\nocite{Register-UK-finds-2020-05-05}
\nocite{Cybernetica-Proposes-2020-05-06}
\nocite{Panoptykon-ProteGo-Safe-2020-05-06}
\nocite{SDZ-Corona-App-2020-05-06}
\nocite{euobserver-Dutch-soap-opera-2020-05-07}
\nocite{MIT-covid-tracing-2020-05-07}
\nocite{Wired-Just-how-2020-05-12}
\nocite{Gizmodo-UKs-Contact-Tracing-2020-05-13}
\nocite{BBC-News-Why-Indias-2020-05-15}
\nocite{POLITICO-States-struggle-2020-05-17}
\nocite{Nature-Can-they-slow-2020-05-19}
\nocite{Guardian-Covidsafe-2020-05-23}
\nocite{Nature-Comment-Ethical-guidelines-2020-06-04}
\nocite{RTL-LU-Lockdowns-averted-2020-06-09}
\nocite{Politico-Privacy-fears-2020-06-04}
\nocite{MIT-Technology-coronavirus-apps-2020-06-05}
\nocite{Top10VPN-Digital-Rights-2020-06-10}
\nocite{LeMonde-StopCovid-2020-06-10}
\nocite{Politico-Google-Apple-2020-06-10}
\nocite{NatureComment-Ten-reasons-2020-05-21}
\nocite{AdaLovelace-Something-to-declare-2020-06-02}
\nocite{EDRi-COVID-Tech-2020-06-10}
\nocite{Guardian-Norway-2020-06-15}
\nocite{BBCNews-Alarm-Kuwait-2020-06-16}
\nocite{EuropeanCommission-interoperability-2020-06-16}
\nocite{Guardian-democracies-2020-06-16}
\nocite{ScientificAmerican-AntibodyTests-2020-06-17}
\nocite{Guardian-UK-abandons-2020-06-18}
\nocite{RTL-German-app-2020-06-25}

\bibliographystyle{plainnat}
\bibliography{covid,wojtek,wojtek-own}

\end{document}